\documentclass[runningheads, anonymous]{llncs}
\usepackage[dvipsnames,table,xcdraw]{xcolor}
\newcommand{\cmark}{\color{ForestGreen} {\ding{51}}}%
\newcommand{\xmark}{\color{BrickRed}{\ding{55}}}%

\usepackage{balance}
\usepackage{todonotes}
\usepackage{xspace}
\usepackage{url}
\usepackage{booktabs}
\usepackage{hyperref}

\providecommand{\vs}{vs.\xspace{}}
\providecommand{\ie}{\emph{i.e.,\xspace{}}}
\providecommand{\eg}{\emph{e.g.,\xspace{}}}

\providecommand{\etc}{\emph{etc.\xspace{}}}

\usepackage{cite}
\usepackage{amsmath,amssymb,amsfonts}
\usepackage{algorithmic}
\usepackage{graphicx}
\usepackage{textcomp}
\usepackage{subcaption}
\usepackage{pifont}   % for \xmark
\usepackage[labelformat=parens,labelsep=quad,skip=3pt]{subcaption}
\begin{document}
\title{SunBlock: Cloudless Protection for IoT Systems}
\author{Vadim Safronov\inst{1} \and
Anna Maria Mandalari\inst{2} \and
Daniel J. Dubois\inst{3} \and
David Choffnes \inst{3} \and
Hamed Haddadi \inst{1}}
\authorrunning{V. Safronov et al.}
\institute{Imperial College London, London, UK 
\email{\{v.safronov,h.haddadi\}@imperial.ac.uk} \and
University College London, London, UK 
\email{a.mandalari@ucl.ac.uk} \and
Northeastern University, Boston, US 
\email{choffnes@ccs.neu.edu, d.dubois@northeastern.edu}
}

%
\maketitle

\begin{abstract}
With an increasing number of Internet of Things (IoT) devices present in homes, there is a rise in the number of potential information leakage channels and their associated security threats and privacy risks. Despite a long history of attacks on IoT devices in unprotected home networks, the problem of accurate, rapid detection and prevention of such attacks remains open. Many existing IoT protection solutions are cloud-based, sometimes ineffective, and might share consumer data with unknown third parties. This paper investigates the potential for effective IoT threat detection locally, on a home router, using AI tools combined with classic rule-based traffic-filtering algorithms. Our results show that with a slight rise of router hardware resources caused by machine learning and traffic filtering logic, a typical home router instrumented with our solution is able to effectively detect risks and protect a typical home IoT network, equaling or outperforming existing popular solutions, without any effects on benign IoT functionality, and without relying on cloud services and third parties. 
    
\end{abstract}

\section{Introduction}

Internet of Things (IoT) devices are increasingly popular, pervasive, and promise a wide range of functionality including remote health monitoring, adaptive climate and lighting, automated control of various appliances and assets within smart spaces. The heterogeneity of IoT device types, functionality, and their applications lead to major security and privacy threats for smart home users. IoT devices are manufactured all over the world, send data globally, may be updated over-the-air,
and use a myriad of third-party software libraries~\cite{ren-imc19}. A single vulnerability in one part of the IoT supply chain can have broad security and privacy impacts for IoT users. For example, compromised IoT devices can compromise user privacy, \eg{} exposing sensitive user activities on personal computers by reading power consumption from a compromised smart plug~\cite{Conti}, among many other serious threats described in recent literature~\cite{Setayeshfar, Karale,Swessi,Sadek}. Also, the combination of IoT device sensors with Internet connectivity poses substantial risks of personally identifiable information (PII) exposure~\cite{ren-imc19}.

To address such risks, several IoT protection solutions have been proposed so far, but they
pose several issues, including their lack of completeness. Evidence from prior research~\cite{Mandalari_IoT_analysis} suggests that popular IoT protection solutions fail to accurately detect and mitigate a significant proportion of threats, particularly those available with basic subscriptions. Moreover, most IoT protection solutions largely rely on cloud-based ML models for detecting attacks. Such cloud-based models necessarily require vendors to collect personal and sensitive data from the networks they protect, and this itself poses the risk of such data being shared with third-parties, used for purposes beyond threat detection, and even being compromised by attacks on cloud infrastructure. In contrast, implementing IoT protection solutions locally can enhance user privacy, obviating the need to disclose personal and potentially sensitive data to other parties outside the local network. Such an edge protection approach can also improve robustness and stability of operations by eliminating the dependence on cloud services. However, a key open question is how feasible such locally run solutions are, and how well they work compared to cloud-based ones.

In this paper, we investigate the possibility of using threat detection and prevention algorithms on a home router, as well as measure quantitative implications on the router’s memory (RAM), processing power (CPU), bandwidth (BW), and whether consuming these resources has any impact on the smooth operation of IoT devices and applications.
Specifically, we demonstrate the feasibility of a completely local IoT protection solution, named SunBlock, which combines existing rule-based traffic filtering tools and machine-learning libraries to provide a resource-efficient prototype combining both intrusion detection~(IDS) and intrusion prevention~(IPS) systems. SunBlock operates on a typical home router, thus improving user privacy without the need
for end users to share potentially sensitive data with cloud-based IDS/IPS solutions as well as bypassing the expense of their premium subscriptions.
We conclude this paper by exposing our SunBlock prototype to a set of various IoT threats, demonstrating that it can detect a spectrum of threats more than twice as large as that detected by commercial IoT protection solutions~\cite{Mandalari_IoT_analysis}. The paper's code is available online~\cite{sunblock}.

\noindent\textbf{Goals.} The paper targets two main research questions (RQs) detailed below.
\noindent \emph{RQ1: How can we replace commercial cloud-based safeguards by a threat detection and prevention software running locally on a home router?}
We address this by designing a new cloudless threat detection/prevention approach (\S{}\ref{sec:design}) and implementing it on popular off-the-shelf router hardware (\S{}\ref{sec:implementation}).\\
\noindent \emph{RQ2: What is the performance of our solution in terms of overhead and threat detection capability?}
We address this by establishing a rigorous evaluation methodology (\S{}\ref{sec:evalmethod}) and performing an extensive evaluation~(\S{}\ref{sec:eval}) to evaluate: (\emph{i}) the overhead of running AI and rule-based IDS locally in terms of router memory, processing power, and bandwidth; (\emph{ii}) how successfully our cloudless solution detects popular attacks compared to existing cloud-based IoT security solutions.

\noindent\textbf{Non-goals.} The following topics are not the focus of this paper.\\
\noindent \textit{Novel ML/AI algorithms for anomaly detection.} We assess existing ML methods for their feasibility to run on constrained devices performing IPS functions.\\
\noindent \textit{Classification of end-user traffic, such as smartphones, tablets and PCs.} The main focus of the paper is on protecting smart home IoT devices.

\section{Threat model}
\label{sub:threat_model}

In this paper we investigate the problem of protecting a typical home network from threats deriving from its IoT devices. 
We focus on instrumenting the home router providing connectivity to the IoT devices with existing algorithms (\ie{} combining IDS/IPS functions with a classic rule-based approach) to investigate the feasibility of running anomaly detection at the edge.
This paper is not about proposing novel ML/AI algorithms or about surveying and comparing different AI/ML models for IoT anomaly detection, as done in previous studies~\cite{netml-paper, dev_ident_3, Kolcun_retraining}.

\begin{figure}[t]
    \centering
    \includegraphics[width=0.7\linewidth]{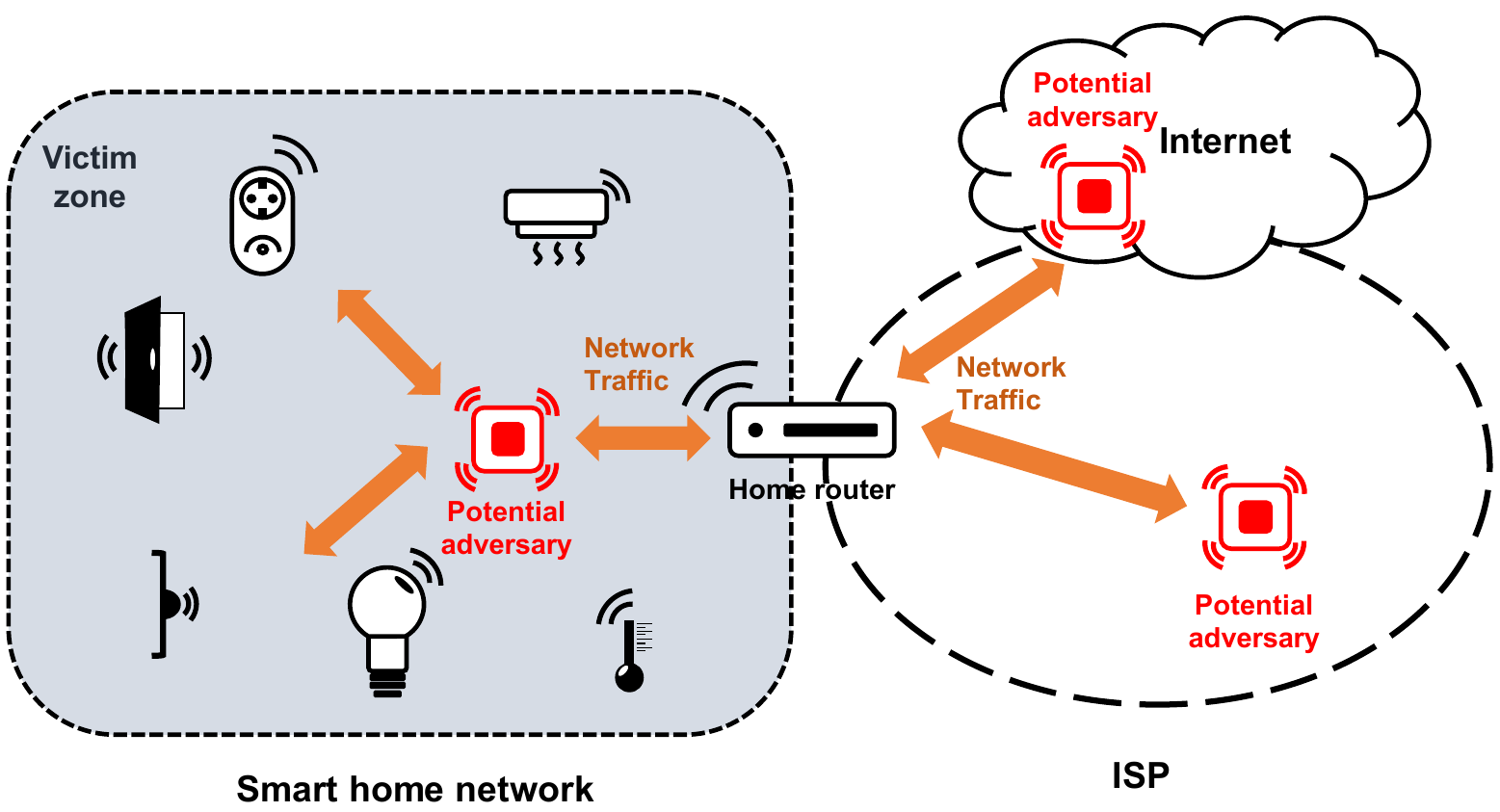}
    \caption{Threat model diagram. Victim zone is a smart home with connected IoT devices. A potential adversary can analyze traffic and conduct privacy/security attacks from smart home, ISP and wider zones, such as the Internet.}
    \label{fig:threat_model}
\end{figure}

Our threat model (Figure~\ref{fig:threat_model}) assumes a typical smart home deployment, with a home router connected to the Internet (WAN interface) creating and providing connectivity to a local smart home network ((W)LAN interface), where IoT devices are deployed.
In this setting, we assume the victim to be a user of IoT devices, and the adversary any entity that can gain network access either from the WAN side (\eg{} an attacker in the ISP network or the Internet) or the (W)LAN side (\eg{} a compromised IoT device).
In both cases the adversary can abuse network access and capture network data (such as traffic rates, protocols used, source/destination IP addresses, \etc{}) 
that can result in security and privacy threats for the victim. 
For example, network data can be used to conduct analyses to infer victim's sensitive data (\eg{} user activities~\cite{ren-imc19}), and assess and exploit vulnerabilities~\cite{paracha2021iotls}. The threats we focus on in this paper are the following:
%hat a router is installed in a smart home with LAN/WLAN-connected IoT devices and the router's WAN port connected to an ISP. From both LAN/WAN sides various entities have potential access to network data (such as traffic rates, protocol used, source/destination IP addresses etc.) from where it is possible to conduct further analysis to infer user activities, privacy-sensitive data and network vulnerabilities. The aforementioned threats are divided into two main groups:

\noindent \textbf{Security:} anomalous traffic, anomalous upload as well as various flooding (\eg{}~DoS) and scanning attacks (e.g. port-scanning).

\noindent \textbf{Privacy:} PII exposure, unencrypted network/application data.

\section{System Design}
\label{sec:design}
\begin{figure}[t]
    \centering
    \includegraphics[width=0.7\linewidth]{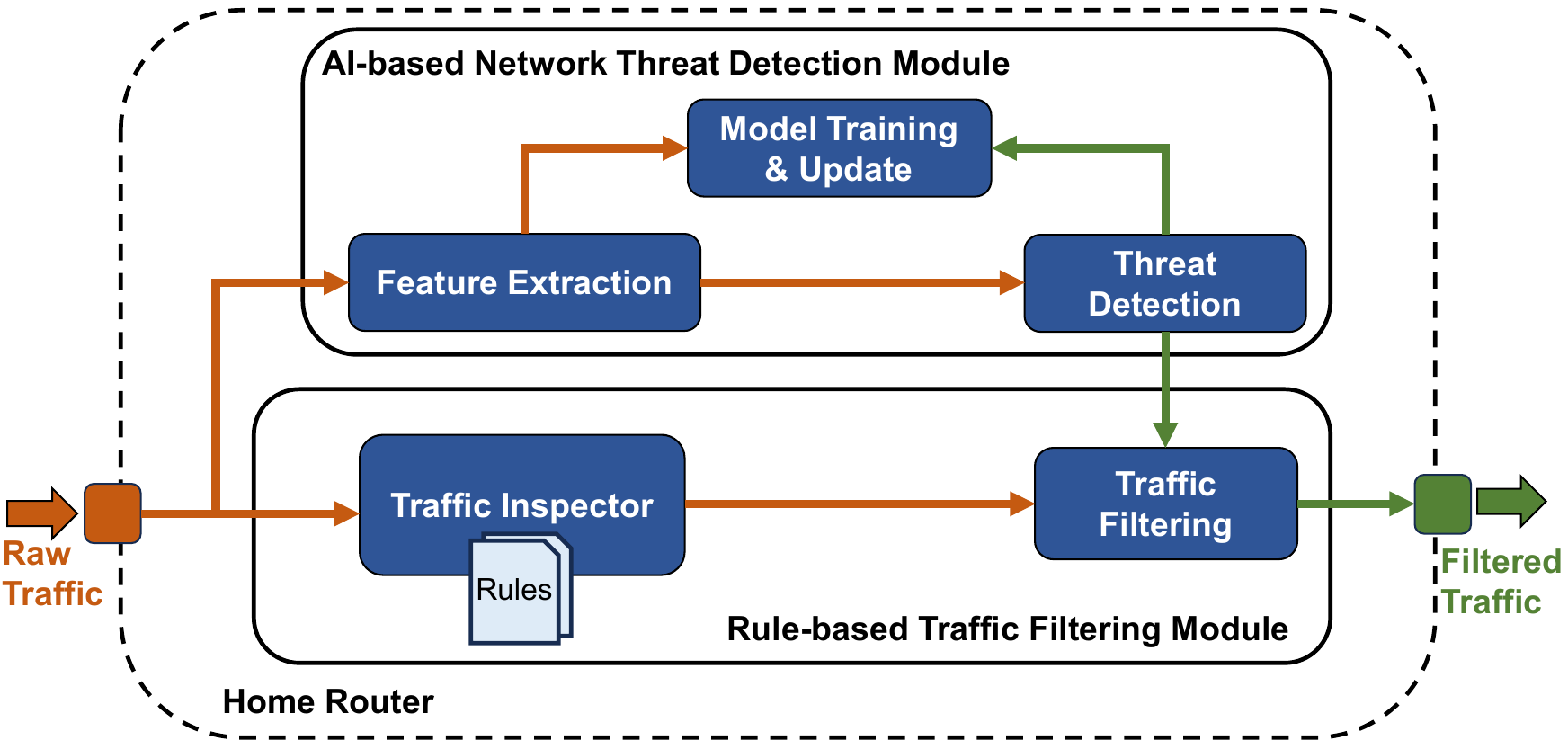}
    \caption{SunBlock architecture running on a home router.}
    \label{fig:achitecture}
\end{figure}

In this section we propose the design of SunBlock, \ie{} an approach for detecting IoT threats at a home router without relying on the cloud. 
Figure~\ref{fig:achitecture} shows the architecture of our system. SunBlock runs on a home router and is composed of: (\emph{i}) a rule-based traffic filtering module; and (\emph{ii}) an AI-based network threat detection module.
Raw IoT traffic traversing the home router is inspected against a defined set of IPS rules~\cite{10.1145/1741906.1741914}, the features are extracted and fed to the trained threat detection model. Both rule-based and AI threat detection steps are performed in parallel and described in detail below.

\subsection{Rule-based Traffic Filtering Module}

This module is responsible for blocking unwanted or potentially harmful traffic while allowing safe and necessary traffic to pass, thereby enhancing the overall security of the network. This module is composed of the following components. 

\noindent \textbf{Traffic Inspector}: As network packets arrive, they are inspected against a defined set of rules at various levels: IP (\ie{} the source and destination IP addresses are checked), transport level (TCP or UDP source and destination ports), and/or at the application level (where the actual content of the packets is inspected, if the HTTP protocol is used, for example).

\noindent \textbf{Traffic Filtering}: Based on the set of pre-defined rules, allowed traffic passes further, while all the inspected traffic matching the blocking rules is dropped.

\subsection{AI-based Network Threat Detection Module}
AI anomaly detection identifies irregular traffic patterns that could indicate potential security threats. 
The module consists of the following components.

\noindent \textbf{Feature Extraction.} Similar to rule-based traffic filtering, the first step includes collecting and processing raw network data, including information about source and destination IP addresses, port numbers, packet sizes, time durations,~\etc{} The collected data is then transformed into a set of features, which are measurable properties or characteristics of the observed network traffic. These could include rate of data transmission, number of connections to a specific server, or frequency of a certain type of packet.

\noindent \textbf{Model Training.} An ML model is trained locally on the extracted dataset of network features. At the training phase, the model learns what `normal' network traffic looks like.

\noindent \textbf{Threat Detection.} SunBlock uses the trained model for comparing live incoming network traffic to the `normal' patterns learned during training. If the live traffic deviates significantly from normal patterns, SunBlock reports a threat.

\noindent \textbf{Model Update.} To support high threat detection accuracy, the model continues to learn and refine its understanding of what constitutes normal behavior.

\section{Implementation}
\label{sec:implementation}
This section answers our first research question by describing a SunBlock prototype, demonstrating that IoT protection can run at the edge, on a home router.  
\subsection{Hardware}
To demonstrate our approach, we use an off-the-shelf middle-range home router: a LinkSys WRT3200ACM~\cite{linksys}. The reason for this choice is its popularity and support for OpenWrt~\cite{openwrt}, a Linux-based OS targeting embedded devices. Due to the router's limited flash memory (256MB), we use a USB stick to allocate 4GB and 512MB partitions for AI libraries/dependencies and swap space, respectively.

\subsection{Rule-based Traffic Filtering}
We use Snort3~\cite{snort3} as a rule-based IDS/IPS. The reason for this choice is the extensive support for rule-based detection techniques, deep packet inspection and various customization options, making it flexibly well-suited for IoT threat prevention use cases. We prefer Snort3 over other solutions like Fail2ban~\cite{fail2ban}, Zeek~\cite{zeek}, and Suricata~\cite{suricata} due to its comprehensive rule-based detection techniques, extensive customization options, and flexibility. Other tools have either narrower focus (\eg{} Fail2Ban focuses on log-based IDS/IPS) or require more scripting overhead to match Snort3's functionality. Besides utilizing the Snort3 community rules~\cite{snort_rules}, we enhanced them by adding blocking logic for DoS, scanning attacks, and unencrypted HTTP traffic.

We configure Snort3 to run in NFQ IPS mode, which allows it to operate as an inline IPS. In this mode, Snort3 intercepts network traffic using Netfilter~\cite{netfilter}, a packet filtering and manipulation framework in the Linux kernel. In NFQ IPS inline mode, Snort3 receives network packets before they reach their target, thus allowing the system to react promptly to any suspicious traffic.

Upon receiving packets, SunBlock performs packet inspection by applying the pre-defined rules serving as criteria for identifying and blocking security and privacy threats. The inspection process involves protocol decoding, traffic analysis, and pattern matching against the defined rule set. Based on the results of the inspection, SunBlock applies rule-based actions. It determines if a packet matches any rules in its rule set and takes actions accordingly, such as dropping the packet or blocking the entire malicious traffic flow. 

\subsection{AI-based Network Threat Detection}
For our threat detection model, we utilize the netml library~\cite{netml-paper}~\cite{netml-code}, a robust tool designed for feature extraction and threat detection in network traffic traces.

While the netml library is primarily intended for powerful machines with ample RAM and CPU resources, we successfully adapt it to be used on home routers. As netml is initially dedicated to x86 architectures, we adapt it for the ARMv7l architecture, commonly used in modern Wi-Fi routers and other embedded networked systems with limited resources. This portability enhances SunBlock's accessibility and flexibility, allowing it to run on a wider range of router configurations. We incorporate netml into a Docker image based on Debian version 10. The resulting image is open-source and accessible online~\cite{sunblock}. After a series of deployment installations and tests, we selected the following versions of netml dependencies, a combination of which allows for the successful running of ML anomaly detection logic on the ARMv7l architecture: numba = 0.56.0, netaddr = 0.8.0, numpy = 1.22.0, pandas = 1.5.3, pyod = 1.0.9, scapy = 2.4.5 and scikit-learn = 0.24.1.

For anomaly detection tasks, SunBlock employs the One-Class Support Vector Machine (OCSVM) model in unsupervised mode with the packet interarrival time (IAT) feature. The OCSVM model is recognized for its outlier detection capabilities, lightweightness and efficacy~\cite{OCSVM_1, OCSVM_2, OCSVM_3}. The precision of the netml-based OCSVM implementation, used by SunBlock, has been extensively evaluated by the netml authors on various datasets, showing high precision for IoT anomaly classification tasks when using the IAT feature~\cite{netml-paper}. The Area Under the ROC Curve (AUC) metric is 0.95 for detecting infected devices, 0.96 for detecting novel devices, and 0.81--0.97 for detecting abnormal device activity. The chosen OCSVM configuration has been wrapped into a real-time threat detection logic having the following characteristics:

\begin{itemize}
    \item Our system is configured to capture batches of 200 packets\footnote{We obtained this number empirically after extensive testing showing a good trade-off between ML accuracy and reaction time.}, thus capturing and classifying network traffic every batch.
    \item Our model is set up to learn from the most recent 7 days of network traffic. This period is chosen based on research indicating a decrease in model accuracy beyond a week without retraining~\cite{Kolcun_retraining}.
    \item Upon detection of a threat, SunBlock uses iptables~\cite{iptables_man} for blocking malicious sources of traffic.
\end{itemize}

\section{Evaluation Methodology}
\label{sec:evalmethod}
In this section we describe how we assess the performance of SunBlock in terms of overhead and threat detection capability. 
We first describe the IoT testbed we use for analyzing SunBlock, then how we emulate the threats, and finally describe the experiment setting and the evaluation metrics.

\subsection{Evaluation Testbed}
\label{sub:testbed}
To assess the efficacy of SunBlock, we build an IoT testbed that emulates a typical smart home setting. According to previous research~\cite{IoT_inspector} that examined smart home traffic from 4,322 households, an average home tends to host around 10 IoT devices. Based on this number, and considering popular IoT device models and categories, we build our testbed using the 10 IoT devices detailed below.

\begin{itemize}
    \item \textbf{Smart speaker:} Amazon Echo Spot, Google Home.
    \item \textbf{Video:} Amazon Fire TV.
    \item \textbf{Camera:} Yi Home Camera, Blink Home Camera.
    \item \textbf{Home automation:} Nest thermostat, TP-Link Kasa plug, WeMo plug, Gosund bulb, TP-Link Kasa bulb.
\end{itemize}

The IoT devices are performing daily activities within a 2.4 GHz WiFi home network in a location similar to a two-bedroom flat. We trigger the IoT device functionality by using a methodology similar to~\cite{mandalari2021blocking} for a period of one week.

\subsection{Threats Emulation}
\label{sec:evalmethod_threat}
For triggering the threats, we employ a Raspberry Pi Model B+ (RPi) as a threat generation node, equipped with 4GB RAM and running a Debian image. 
The RPi is connected to the (W)LAN side of the home network along with the IoT devices. 
In order to check the SunBlock performance in realistic scenarios, the RPi used real attack scripts from an open GitHub repository~\cite{safeguards_study_git} containing threat emulation scripts from recent research on the efficiency of IoT protection systems~\cite{Mandalari_IoT_analysis}.
The threats we emulate are the following.

\noindent \textbf{Anomalous Traffic.} RPi acts on behalf of the Echo Spot device by spoofing its IP and MAC addresses. However, instead of sending typical Echo Spot traffic, the RPi injects previously captured traffic from the Google Home. We inject this traffic using tcpreplay~\cite{Tcpreplay2023}, where we replace the IP and MAC addresses of the Google Home with Echo Spot's ones.

\noindent \textbf{Anomalous Upload.} We emulate anomalous upload behavior by making the RPi upload a video to an external unauthorized server while spoofing the IP and MAC addresses of the Yi home camera using the tcpreplay tool~\cite{Tcpreplay2023}.

\noindent \textbf{DoS Flooding.} For SYN, UDP, DNS and HTTP flooding threats, the RPi uses suits of attacks from Kali Linux emulating a typical Mirai-infected device~\cite{mirai_botnet}. 

\noindent \textbf{Privacy script.} The RPi acts on behalf of Google Home (by spoofing its IP and MAC addresses) and transmits unencrypted sensitive data, such as login details, passwords and other private data, to an external server through HTTP.

\noindent \textbf{Scanning attacks.} We emulate OS and port scanning using Nmap~\cite{lyon2008nmap}. This information can be utilized by a potential attacker to determine the most effective attack strategy for the detected OS, taking advantage of the open ports.

\subsection{Experiments Description and Evaluation Metrics}

Prior to executing the threat emulation scripts, we benchmark the resource utilization (CPU, RAM, and bandwidth) of the home router running SunBlock with respect to a router not running it.
We continuously monitor these metrics using a Python-based script, which captures resource consumption data over a span of 3.5 hours of regular IoT traffic.

To evaluate SunBlock performance, we time the training process under four distinct conditions: both modules operational, only the AI-based module active, only the Rule-based module active, and an unprotected scenario. Under each condition, the model underwent five training iterations. A background script records CPU usage, RAM (actual RAM used by the system, excluding buffers/cache), bandwidth consumption, and the start and end times of each training instance.

We conduct each attack for 100 seconds in 10 separate iterations. After each attack, we reset the network to its normal operational state before we start the next iteration. The RPi logs both the start and end times for each individual attack.
%, providing a time-stamped record of each event. 
We conduct all experiments in a home network environment equipped with the SunBlock router. We then repeat the same set of experiments in a network with the same router, but with SunBlock disabled.

To measure resource utilization of the home router, both with and without SunBlock, we execute a background Python-based script during each attack, which records the utilization of \emph{CPU, RAM, and BW} every second.

To measure \emph{threat prevention latency}, we execute a script to capture SunBlock's notifications, which provided timestamps and types of blocked threats. We calculate the prevention latency as the time between the start of the attack and the arrival of a SunBlock's notification indicating a blocked threat.

\section{Evaluation}
\label{sec:eval}

In this section we answer our second research question by evaluating the performance of SunBlock in terms of overhead and threat detection capability.

\subsection{Performance Overheads}

Table ~\ref{tab:eval_training} shows model training performance overhead in distinct conditions: fully protected (Rule-based and AI-based enabled modes), partially protected (either Rule-based or AI-based mode is enabled), and unprotected (disabled SunBlock). 
\begin{table}[h!]
\centering
\caption{ML training time and resource usage for various protection levels.}
\label{tab:eval_training}
\resizebox{0.65\linewidth}{!}{
\renewcommand{\arraystretch}{1.5} % Adjust this value as necessary
\begin{tabular}{p{0.30\columnwidth}||p{0.1\columnwidth}||p{0.1\columnwidth}||p{0.12\columnwidth}||p{0.2\columnwidth}}
\textbf{Protection Level} & \textbf{CPU \newline(\%)} & \textbf{RAM (MB)} & \textbf{swap \newline (MB)} & \textbf{Training Time (s)} \\ 
\hline\hline
Rule-based \& AI-based & 18 $\pm$3 & 444 $\pm$4 & 296 $\pm$21 & 924 $\pm$253 \\
\hline
AI-based only & 26 $\pm$2 & 442 $\pm$6 & 197 $\pm$28 & 429 $\pm$171 \\
\hline
Rule-based only & 32 $\pm$4 & 423 $\pm$9 & 132 $\pm$20 & 180 $\pm$22 \\
\hline
Unprotected & 39 $\pm$2 & 410 $\pm$3 & 55 $\pm$1 & 113 $\pm$10 \\
\hline\hline
\end{tabular}
}
\end{table}

When deployed in tandem, Rule-based and AI-based modules constitute the most secure mode. Although the fully protected mode takes the longest time for training --- up to 20 minutes --- it has the lowest CPU utilization among all modes. Compared to AI-based/Rule-based only modes, the fully protected mode has increased training time by a factor of 1.5--6.5, and up to a factor of 10.5 when compared to the unprotected mode. However, the duration of fully protected training remains within an acceptable range and can be scheduled during off-peak hours to minimize disruptions. 

Memory usage is also critical for a home router. During the fully protected mode, swap memory usage peaks at around 300MB, which is manageable for a typical home router equipped with sufficient flash memory (either internally or through an external USB stick). It is noticeable that swap is required only for model training and is not used in SunBlock's daily behavior. Swap should be allocated as a separate partition for its isolation from router's storage memory.

\begin{figure}[h]
    \centering
    \includegraphics[width=0.6\linewidth]{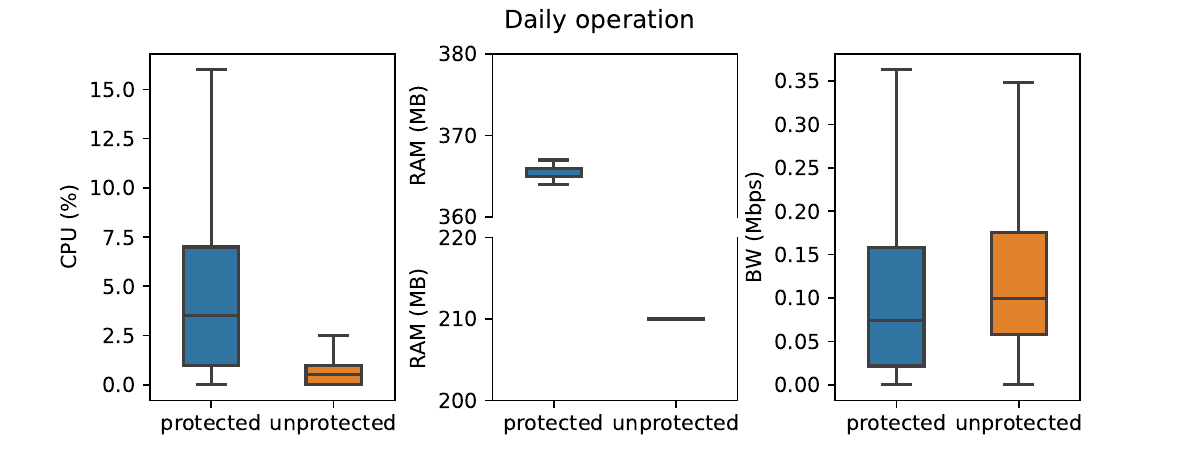}
    \caption{Resource usage in protected \vs{} unprotected mode for regular IoT traffic.}
    \label{fig:idle_performance}
\end{figure}

Figure~\ref{fig:idle_performance} provides a performance comparison between the router in SunBlock's protected mode and the same router in the unprotected mode, during their routine operation without network threats. SunBlock exhibits a wider CPU utilization range compared to the unprotected router; however, the utilization peaks at only $\sim$15\%. The median CPU utilization remains at an approximate 3\%, thus maintaining a negligible impact on overall performance. SunBlock has 1.7 times higher RAM consumption than the unprotected router. While this increment is notable, the overall RAM usage remains well within the router's capacity, leaving about 30\% of free RAM. The bandwidth usage is the same in both modes due to the same traffic conditions under which both protected and unprotected modes were tested.

\begin{figure}[h]
    \centering
    \includegraphics[width=0.6\linewidth]{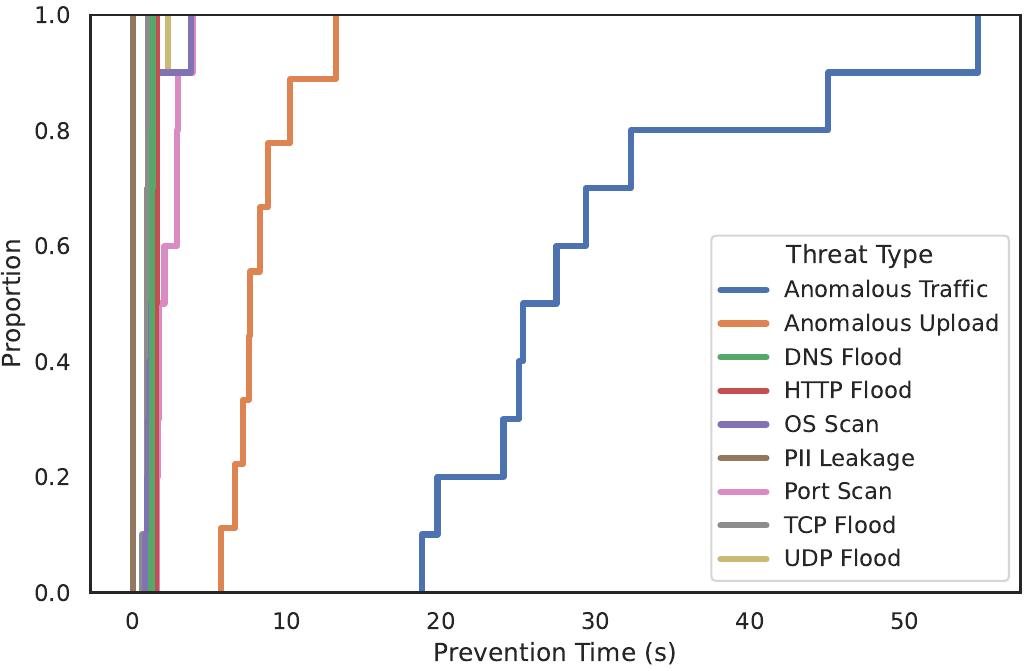}
    \caption{ECDFs of SunBlock's prevention time per each attack.}
    \label{fig:reaction_time}
\end{figure}

Figure~\ref{fig:reaction_time} illustrates the Empirical Cumulative Distribution Functions (ECDFs) representing the times taken by SunBlock to prevent threat types listed in \S{}\ref{sec:evalmethod_threat}. It can be observed that most threats are blocked within $\sim$5 seconds. However, SunBlock requires more time to detect anomaly threats, namely $\sim$10 seconds for anomalous upload and $\sim$50 seconds for anomalous traffic.

\begin{figure*}[h!]
  \begin{subfigure}{6cm}
    \centering\includegraphics[width=\linewidth]{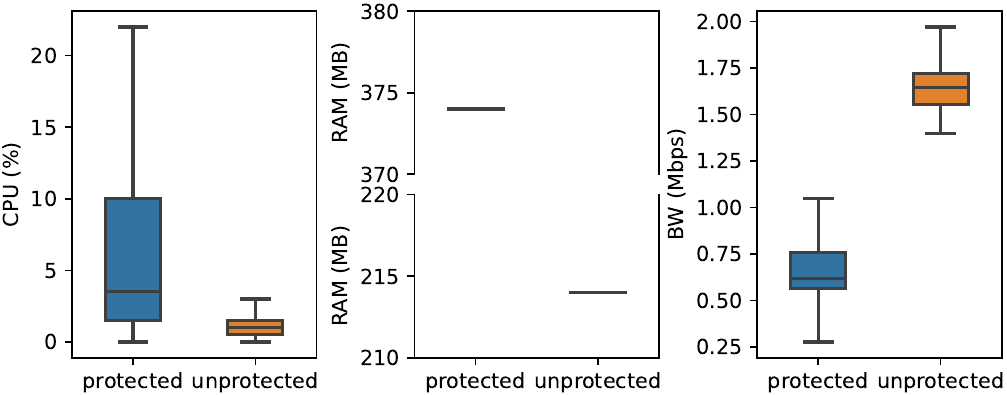}
    \caption{DNS flood}
    \label{fig:dns_flood}
  \end{subfigure}
  \begin{subfigure}{6cm}
    \centering\includegraphics[width=\linewidth]{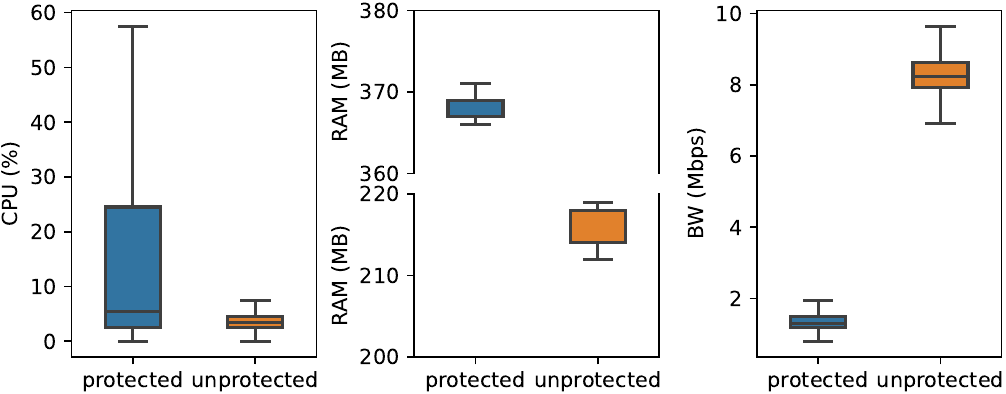}
    \caption{HTTP flood}
    \label{fig:http_flood}
  \end{subfigure}
  \begin{subfigure}{6cm}
    \centering\includegraphics[width=\linewidth]{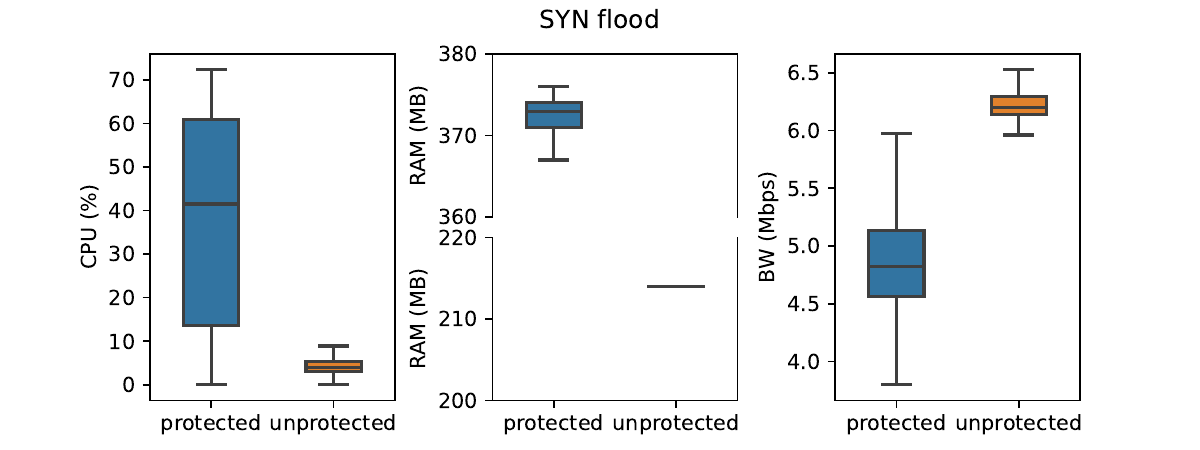}
    \caption{SYN flood}
    \label{fig:syn_flood}
  \end{subfigure}
  \begin{subfigure}{6cm}
    \centering\includegraphics[width=\linewidth]{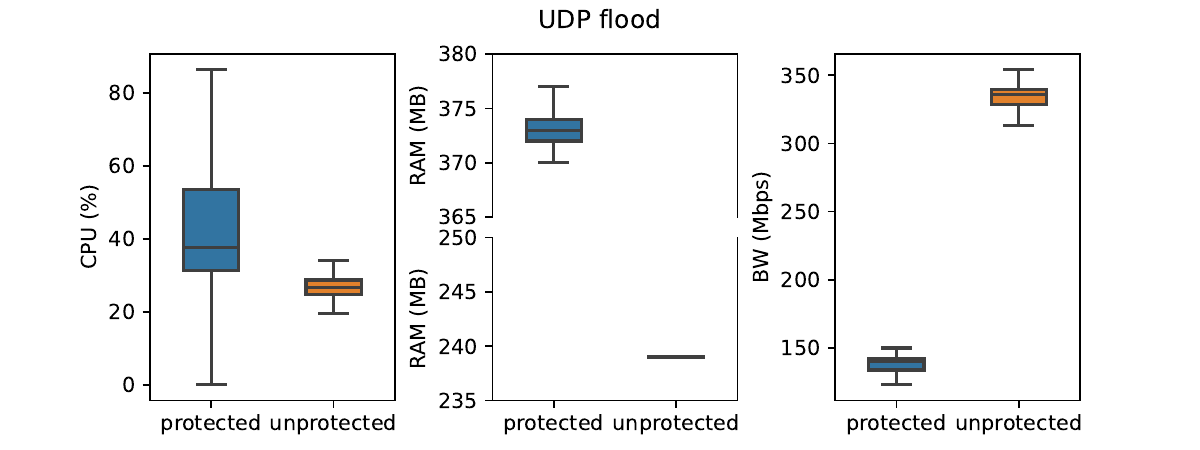}
    \caption{UDP flood}
    \label{fig:udp_flood}
  \end{subfigure}
  \caption{Resource usage in protected \vs{} unprotected mode under DoS attacks.}
\label{fig:flood}
\end{figure*}

\begin{figure}[h!]
    \centering\includegraphics[width=0.6\linewidth]{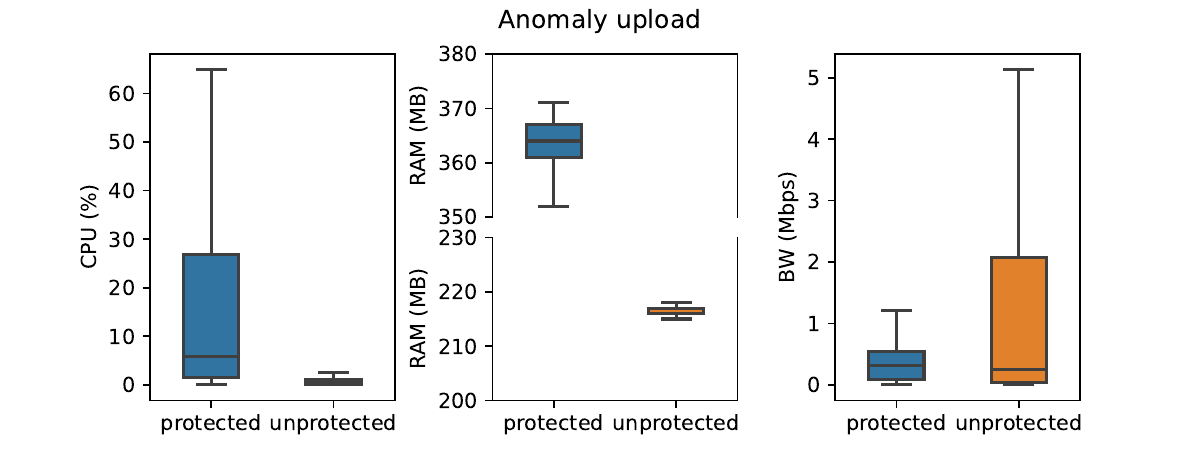}
    \caption{Resource usage in protected \vs{} unprotected mode for anomalous upload.}
    \label{fig:anomal_upload}
  \end{figure}

Figures~\ref{fig:flood}, \ref{fig:anomal_upload} show RAM, CPU and BW overhead of SunBlock under a number of traffic-heavy attacks (DNS, HTTP, SYN, UDP flooding and anomalous upload), compared to an unprotected mode. 

As anticipated, SunBlock exhibits an elevated CPU/RAM usage due to the operational load imposed by Rule-based and AI-based modules actively performing threat mitigation tasks. SunBlock's median CPU consumption varies between 3--40\%, while the RAM usage remains fairly stable within a range of 365--375MB. Conversely, an unprotected router exhibits significantly lower RAM/CPU usage, $\sim$220MB of RAM and 1--25\% of median CPU utilization.

The protective effects of SunBlock are evident in the bandwidth consumption under heavy DoS attacks. The box plots indicate a substantial bandwidth reduction of up to $\sim$8 times during intense DoS attacks, with the strongest reductions during HTTP and UDP flooding. During the attacks, SunBlock maintains an adequate buffer of resources, with  50\% of CPU and 30\% of RAM always available.

\subsection{Threat Detection Capability}

%This subsection discusses the evaluation of the threat detection capability with respect to existing IoT protection solutions.
Table~\ref{tab:eval_comparison} presents a comprehensive comparison of SunBlock's threat detection capability with existing IoT protection solutions. The data on threat detection capabilities of existing solutions is sourced from prior comparative research~\cite{Mandalari_IoT_analysis}, which evaluated a number of most popular IoT protection solutions (Avira~\cite{avira}, Bitdefender~\cite{bitdefender}, F-Secure~\cite{fsecure}, Fingbox~\cite{fing}, Firewalla~\cite{firewalla}, McAfee~\cite{mcafee}, RATtrap~\cite{rattrap}, and TrendMicro~\cite{trendmicro}) using the same threat generation scripts as our evaluation. For the sake of brevity, the table shows merged IoT threat detection results of these protection systems. A check mark against a threat means at least one system can detect it. Results show that SunBlock outperforms IoT protection solutions in terms of threat detection capabilities. 
\begin{table}[h!]
\centering
\caption{Threats detected by existing IoT solutions \vs{} detected by SunBlock.}
\label{tab:eval_comparison}
\resizebox{0.65\linewidth}{!}{
\begin{tabular}{m{0.33\columnwidth}||m{0.20\columnwidth}||m{0.15\columnwidth}}
\textbf{Threat} & \textbf{IoT Protection Systems} & \textbf{SunBlock} \\ 
\hline\hline
Anomalous Traffic & \xmark & \cmark \\
\hline
Anomalous Upload & \xmark & \cmark \\
\hline
SYN Flooding  & \cmark & \cmark \\
\hline
UDP Flooding  & \xmark & \cmark \\
\hline
DNS Flooding & \xmark & \cmark \\
\hline
HTTP Flooding & \cmark & \cmark \\
\hline
Port Scanning & \cmark & \cmark \\
\hline
OS Scanning & \cmark & \cmark \\
\hline
PII Leakage & \xmark & \cmark \\
\hline\hline
\end{tabular}
}
\end{table}

\section{Limitations and Future Work}
\label{sec:limitations}

\noindent \emph{Hardware.} Our approach was tested on an off-the-shelf router with 512MB of RAM. We choose this since full-featured routers offered by the most popular US ISPs have more memory than that (\eg{} the Xfinity xFi XB6 from Comcast has 768MB~\cite{comcast-xfi-xb6} and the CR1000A~\cite{verizon-cr1000a} from Verizon has 2GB). However, some older routers and lower-tier router offerings may still have a lower amount of RAM. In such case, our approach would be slower due to requiring swap space. Future improvements involve developing a more efficient Docker image for ML anomaly detection by utilizing lighter Linux distributions such as Arch Linux, which require more custom configuration compared to Debian-based images. Additionally, it is worthwhile to monitor emerging ML/AI methods and associated compression and pruning techniques that may reduce RAM and CPU usage while maintaining high classification performance.

\noindent \emph{Zero-day threat protection.} Zero-day threats pose a common challenge for ML systems, as learning models require time to gather relevant training data during their first run. Shortening the initial training period can reduce exposure but might compromise accuracy, necessitating more user input for custom classification. This is not always feasible as home users are not typically security experts. SunBlock's Rule-based Traffic Filtering, functioning from day one, reduces the risk of zero-day exposure by protecting against most severe threats, such as flooding and scanning. Additionally, other security measures during the initial setup may include factory resetting all present IoT devices, ensuring they are updated with the latest software versions, and, if feasible, placing them on a separate network segment with limited Internet access.

\noindent \emph{Model Updates.} Since our approach is cloudless, it does not include external (cloud-based) means of updates. The capability of our approach to learn and self-adapt partially address this limitation; however, faster updates can be achieved by allowing the periodic download of updated Snort community rules~\cite{snort_rules} and/or by building and sharing AI models in a privacy-preserving crowdsourced way (\eg{} as discussed in~\cite{briggs2021review,zhou2020privacy}).

\noindent \emph{Configuration.} As discussed before, default parameters of our approach were chosen empirically for our prototype. However, depending on the particularities of deployment scenario and user needs, such parameters may need to be tuned to achieve the desired trade-off between false negative and false positive threat reporting. This limitation can be addressed, for example, by using a interactive ML approaches~\cite{wu2022survey,dudley2018review}, where the user is asked to confirm what to do for each threat until the system learns how to make this decision on its own.

\section{Related Work}

The growing privacy and security issues in the consumer IoT domain have prompted the creation of various tools~\cite{SPIN, mandalari2021blocking} designed to block malicious traffic originating from IoT devices. Despite the promising advances these tools represent, there remains an area for their potential improvement by incorporating AI for the detection of threats. 

Other attempts~\cite{Patel, Dua, palmese2022feature} to shift IoT protection to the edge have limitations. Implementations of~\cite{Patel, Dua} rely solely on classic rule-based traffic filtering, which is less sensitive to anomalous traffic threats and requires additional AI-based classification. The evaluations in~\cite{Patel, Dua, palmese2022feature} were conducted on non-router-specific superior hardware and may not reflect consumer router performance.

There have been recent introductions of AI-based methodologies dedicated to device identification tasks at the edge~\cite{Dev_ident_1, Dev_ident_2, dev_ident_3, dev_ident_4}. However, they alone are insufficient for comprehensive protection against major IoT threats and would benefit from rule-based and AI-based threat detection techniques.

Numerous studies have examined IoT threats, assessing smart home vulnerabilities, attack scenarios, and defense strategies~\cite{IoT_sec_1, IoT_sec_2, IoT_sec_3}. Recent research~\cite{Mandalari_IoT_analysis} introduced a methodology to assess the efficacy of current IoT protection solutions in responding to prevalent security and privacy threats. The associated threat emulation scripts and benchmarks were adopted in our study for testing the effectiveness of our prototype against popular IoT protection solutions.

\section{Discussion \& Conclusion}

The growing ubiquity of IoT devices in residential settings is transforming our homes into sophisticated, interconnected digital environments. As such, the importance of robust and proactive protection measures for these devices is vital, ensuring not only their functionality but also the privacy and security of the vast amounts of personal data they handle. 

Our research reveals that IoT threats can be rapidly mitigated on a home router, equipped with ML/AI anomaly detection and rule-based traffic filtering algorithms. Most types of threats are promptly identified and blocked within the first 5 seconds. Nonetheless, certain threats such as anomalous traffic and anomalous upload require a longer time due to their initial similarity with normal network behavior of other IoT devices in the network. The presented local threat detection approach eliminates the need for dependence on cloud-based IoT security solutions and thus blocks extra channels of potential PII and other user-sensitive data exposure. It is important to note that integrating ML with rule-based traffic analysis in our off-the-shelf router (which has inferior hardware with respect to full-featured routers currently offered by top ISPs in the US~\cite{comcast-xfi-xb6,verizon-cr1000a}) cause a slight increase in CPU utilization (up to $\sim$15\% with an average use of $\sim$3\%) and RAM consumption (by $\sim$150MB) during normal operation, without affecting main router functions. Our plans for further enhancements involve intensive beta testing and precise performance benchmarking against existing cloud-based security solutions.

\noindent\textbf{Acknowledgements.} We thank the anonymous reviewers and our shepherd Roland van Rijswijk-Deij for their constructive and insightful feedback. This work was supported by the EPSRC Open Plus Fellowship (EP/W005271/1), the EPSRC PETRAS grant (EP/S035362/1), and the
NSF ProperData award (SaTC-1955227).

{
  \bibliographystyle{splncs04}
  \bibliography{paper}
}

\end{document}